\documentclass[12pt]{iopart}

\usepackage{graphicx}
\usepackage{iopams}  

\usepackage{subfigure}

\def\be{\begin{equation}}
\def\ee{\end{equation}}
\def\ba{\begin{eqnarray}}
\def\ea{\end{eqnarray}}

\begin{document}

\title[Joint spin and Fortuin-Kasteleyn observables]{Critical
  properties of joint spin and Fortuin-Kasteleyn observables in the
  two-dimensional Potts model}

\author{Romain Vasseur$^{1,2,4}$ and Jesper Lykke Jacobsen$^{2,3,4}$}
\address{${}^1$Institut de Physique Th\'eorique, CEA Saclay,
91191 Gif Sur Yvette, France}
\address{${}^2$LPTENS, \'Ecole Normale Sup\'erieure, 24 rue Lhomond, 75231 Paris, France}
\address{${}^3$Universit\'e Pierre et Marie Curie, 4 place Jussieu, 75252 Paris, France}
\address{${}^4$Institut Henri Poincar\'e, 11 rue Pierre et Marie Curie,
  75231 Paris, France}

\eads{\mailto{romain.vasseur@cea.fr}, 
      \mailto{jesper.jacobsen@ens.fr}}

\begin{abstract}

  The two-dimensional Potts model can be studied either in terms of
  the original $Q$-component spins, or in the geometrical
  reformulation via Fortuin-Kasteleyn (FK) clusters. While the FK
  representation makes sense for arbitrary {\em real} values of $Q$ by
  construction, it was only shown very recently that the spin
  representation can be promoted to the same level of generality. In
  this paper we show how to define the Potts model in terms of
  observables that simultaneously keep track of the spin and FK
  degrees of freedom. This is first done algebraically in terms of a
  transfer matrix that couples three different representations of a
  partition algebra. Using this, one can study correlation functions
  involving any given number of propagating spin clusters with
  prescribed colours, each of which contains any given number of
  distinct FK clusters. For $0 \le Q \le 4$ the corresponding critical
  exponents are all of the Kac form $h_{r,s}$, with integer indices
  $r,s$ that we determine exactly both in the bulk and in the boundary
  versions of the problem.  In particular, we find that the set of 
  points where an FK cluster touches the hull of its surrounding spin 
  cluster has fractal dimension $d_{2,1} = 2 - 2 h_{2,1}$. If one constrains 
  this set to points where the neighbouring spin cluster extends to
  infinity, we show that the dimension becomes $d_{1,3} = 2 - 2 h_{1,3}$.  
  Our results are supported by extensive transfer matrix and Monte Carlo
  computations.

\end{abstract}

\pacs{64.60.De 64.60.F- 05.50+q}
% \maketitle

\section{Introduction}

The two-dimensional $Q$-state Potts model \cite{Potts52} is one of the
most well-studied models in the realm of statistical physics. It can
be addressed on the lattice using traditional approaches \cite{Wu82}
(duality, series expansions, exact transformations) and algebraic
techniques \cite{Martin} (Temperley-Lieb algebra, quantum groups). In
some cases it can be exactly solved using quantum inverse scattering
methods \cite{Baxter82}. When $0 \le Q \le 4$ it gives rise to a
critical theory in the continuum limit, which can in turn be
investigated by the methods of field theory \cite{JJreview} (Coulomb
gas, conformal field theory) or probability theory \cite{BBreview}
(Stochastic Loewner Evolution). Despite of this well-equipped toolbox
and sixty years of constant investigations, the Potts model remains at
the forefront of current research, most recently in the context of
logarithmic conformal field theory \cite{Vasseur1,Vasseur2}.

The partition function of the Potts model reads
\be
 Z = \sum_{\sigma} \prod_{(ij) \in E}
 \exp \left( K \delta_{\sigma_i,\sigma_j} \right) \,,
 \label{Potts}
\ee
where $K$ is the coupling between spins $\sigma_i = 1,2,\ldots,Q$
along the edges $E$ of some lattice ${\cal L}$.  The Kronecker delta
function $\delta_{\sigma_i,\sigma_j}$ equals 1 if $\sigma_i =
\sigma_j$, and 0 otherwise.  The precise choice of ${\cal L}$ does not
affect universal critical properties, so for convenience we choose the
square lattice. The transition line---which gives rise to a critical
theory for $0 \le Q \le 4$---is then given by the selfduality
criterion ${\rm e}^K = 1 + \sqrt{Q}$.

There are two obvious ways to rewrite the local Boltzmann weights:
\be
 \exp \left( K \delta_{\sigma_i,\sigma_j} \right) = \left \lbrace
 \begin{array}{ll}
 w \left[(1 - \delta_{\sigma_i,\sigma_j})w^{-1} +  \delta_{\sigma_i,\sigma_j} \right] \,, \quad &
 w = {\rm e}^K \,, \\
 1 + v \delta_{\sigma_i,\sigma_j} \,, &
 v = {\rm e}^K-1 \,,
 \end{array} \right.
 \label{Boltzmann}
\ee
where we have used that $\delta_{\sigma_i,\sigma_j} = 0$ or $1$.
The partition function (\ref{Potts}) is obtained by expanding
a product of such factors; each term in the expansion corresponds
to choosing, for each $(ij) \in E$, either the first or the
second term in (\ref{Boltzmann}).

Consider first the expansion in powers of $w^{-1}$. Each factor of
$w^{-1}$ comes with the term $(1-\delta_{\sigma_i,\sigma_j})$ that
forces the spins $\sigma_i$ and $\sigma_j$ to take different values;
there is thus a piece of domain wall (DW) on the edge $(ij)^*$ dual to
$(ij)$. Making this choice for each $(ij) \in E$ defines a DW
configuration on the dual lattice ${\cal L}^*$. This DW configuration
defines a graph $G$ (not necessarily connected), the faces of which
are the clusters of aligned spins.  Since we do not specify the colour
of each of these clusters, a DW configuration has to be weighted by
the chromatic polynomial $\chi_{G^*}(Q)$ of the dual graph $G^*$.
Initially $\chi_{G^*}(Q)$ is defined as the number of colourings of
the vertices of the graph $G^*$, using colours $\{1,2,\ldots,Q\}$,
with the constraint that neighbouring vertices have different colours.
This is indeed a polynomial in $Q$ for any $G$, and so can be
evaluated for any real $Q$ (but $\chi_{G^*}(Q)$ is integer only when
$Q$ is integer).  The partition function (\ref{Potts}) can thus be
written as a sum over all possible DW configurations
\be
 Z_{\rm spin} = w^N \sum_{G} \, w^{-{\rm length}(G)}
  \, \chi_{G^*}(Q) \,,
 \label{PottsDW}
\ee
where $N$ is the number of spins, and ${\rm length}(G)$ denotes the
total length of the domain walls.

Alternatively we may consider the expansion in powers of $v$.  The set
of edges $(ij) \in E$ for which the term $v
\delta_{\sigma_i,\sigma_j}$ is chosen defines a graph $H$, the
connected components of which are known as Fortuin-Kasteleyn (FK)
clusters \cite{FK72}. Since spins belonging to the same FK cluster are
obviously aligned, we can perform the sum over $\{\sigma_i\}$ in
(\ref{Potts}) to obtain
\be
 Z_{\rm FK} = \sum_{H} Q^{{\rm \sharp clusters}(H)} v^{{\rm \sharp edges}(H)} \,.
 \label{PottsFK}
\ee
We stress that spins belonging to different FK clusters can still be
aligned (with probability $1/Q$), even if the two clusters are
adjacent on the lattice.

Despite of the fact that $Z_{\rm spin} = Z_{\rm FK} = Z$ exactly, the
different geometrical degrees of freedom ($G$ resp.\ $H$) underlying
the spin and FK formulations give rise to different critical exponents
in the continuum limit for $0 \le Q \le 4$. In both cases it is
possible to build $Z$ from a transfer matrix that acts on respectively
the FK \cite{Blote82} and spin \cite{Dubail10} degrees of
freedom. The transfer matrix is an essential ingredient in both
analytical and numerical studies, and it gives access to the rich
algebraic structures underlying the Potts model. In the FK case this
structure is the Temperley-Lieb algebra \cite{Martin}, whereas the
particular partition algebra that emerges in the spin case
\cite{Dubail10} still awaits a complete study.

Critical exponents corresponding to FK clusters were first derived by
Coulomb gas techniques (see e.g.\ \cite{JJreview}) and have
subsequently been confirmed by the rigorous methods of SLE (see e.g.\
\cite{BBreview}). In particular the probability that $\ell$ distinct
FK clusters propagate between small neighbourhoods of two distant
points defines a critical exponent that can be found in the Kac table
of CFT as $h_{0,\ell}$ in the bulk case, and as $h_{1,1+2\ell}$ in the
boundary case \cite{DS_NPB87}. The spin case turns out to be richer,
since one needs in addition to specify the respective colours of the
propagating spin clusters. A pair of adjacent clusters having
different (resp.\ identical) colours give rise to a thin (resp.\ a
thick) DW. The exponents corresponding to $\ell_1$ thin DW and
$\ell_2$ thick DW can be identified as $h_{\ell_1-\ell_2,2\ell_1}$ in
the bulk case and as $h_{1+2(\ell_1-\ell_2),1+4\ell_1}$ in the
boundary case \cite{Dubail10}.  At present this result is based on
(substantial) numerical evidence, supported by analytical arguments
for the special case $\ell_1 = 0$.

The purpose of this paper is to study the {\em joint} properties of FK
and spin clusters, by simultaneously keeping track of the geometrical
degrees of freedom $G$ and $H$ appearing in
(\ref{PottsDW})--(\ref{PottsFK}). The corresponding geometrical
observables are defined precisely in section~\ref{sec:obs}. We show
that making certain physical assumptions---and insisting on recovering
the results for FK \cite{DS_NPB87} and spin \cite{Dubail10} clusters
as special cases---leads to a conjecture for the scaling dimensions of
joint observables in the general case.  This conjecture generalises
all the preceeding results and gives access to new information on the
interaction between FK and spin clusters.  All the scaling dimensions
fit into the Kac table $h_{r,s}$ with integer indices $r$ and $s$.

This result constitutes a further step in the programme \cite{CBL} of
classifying non-unitary boundary conditions in two-dimensional
geometrical models.

In section~\ref{sec:TM} we define a transfer matrix that contains
complete information about these joint spin-FK observables. It
formally acts on three coupled partition algebras. We study its
spectrum numerically in section~\ref{sec:diag}, finding strong support
for our general conjecture for the scaling dimensions of geometrical
observables.  In section~\ref{sec:MC} we further corroborate the
conjecture by studying the most relevant among the new exponents by
Monte Carlo simulations. We discuss our findings in
section~\ref{sec:disc}.

\section{Joint spin and FK observables}
\label{sec:obs}

We begin by reviewing the definition of observables in the spin
representation of the Potts model \cite{Dubail10}. For the purposes of
this discussion it suffices to consider two-point correlation
functions. We then ask how the probability, that a certain number
$\ell$ of spin clusters---each corresponding to a given value of the
spin $\sigma_i$---connect a small (in units of the lattice spacing)
neighbourhood $A$ to another small neighbourhood $B$, decays when the
distance $x$ between $A$ and $B$ increases. It is convenient to
specify the two-point function by just giving the $\ell$ colour labels
$\sigma_i$. By the $S_Q$ permutation symmetry only the relative
colours matter. For instance, with $\ell=4$ the labels $1123$ and
$2241$ specify the same correlation function.

Alternatively one may think of this question in terms of the DW that
separate pairs of adjacent (when moving along the rim of $A$ or $B$)
spin clusters. Each DW can be either {\em thin} or {\em thick},
depending on whether the two adjacent clusters have different or
identical colours. The reason for this nomenclature is that adjacent
spin clusters with different colours can touch (in which case the
corresponding DW has a width of one lattice spacing), whereas if the
colours are identical they cannot (since then there would be no
separating DW). In other words, the width of a thick DW is at least
two lattice spacings. The results of \cite{Dubail10} show that this
distinction is important, since it survives in the continuum limit
and gives rise to different critical exponents.

\begin{figure}[t]
\begin{center}
    \includegraphics[width=8.0cm]{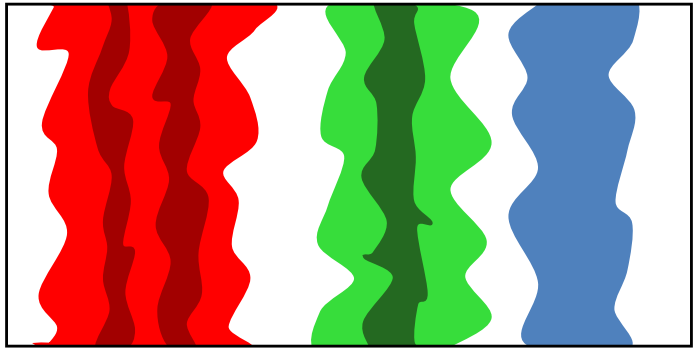}
  \caption{Schematic example of excitations considered in this paper, with three spin clusters 
carrying different coulours propagating along the imaginary time direction. The shaded regions represent 
propagating FK clusters living inside the spin clusters.
  Using the notations~(\ref{gen_label}), this excitation corresponds to $1^2 2^1 3^0$.}
  \label{fig1}
\end{center}
\end{figure}

An obvious but important consequence of (\ref{Boltzmann}) is that if
two points belong to the same FK cluster, it implies that they also
belong to the same spin cluster. The opposite implication is however
not true. This means that FK clusters live ``inside'' spin clusters,
and in particular a given spin cluster can contain several distinct FK
clusters. This suggests introducing a more general two-point function
where each propagating spin cluster contains a given non-negative
number of propagating FK clusters. We describe this by a label
\be
 \sigma_1^{\alpha_1} \sigma_2^{\alpha_2} \ldots \sigma_{\ell}^{\alpha_\ell} \,,
 \label{gen_label}
\ee
where $\sigma_k$ is the colour of $k$'th spin cluster, and $\alpha_k =
0,1,2,\ldots$ is the number of FK clusters living inside the $k$'th
spin cluster. An example of such an excitation  is shown in Fig.~\ref{fig1}.

\subsection{Result for bulk and boundary critical exponents}

Two different geometries are of interest. In the bulk case, the
neighbourhoods $A$ and $B$ are at arbitrary, but widely separated,
locations in the infinite plane. This is conformally equivalent to an
infinitely long cylinder---a strip with periodic boundary
conditions---with $A$ and $B$ situated at the two extremities. In the
boundary case, the geometry is that of the upper half plane, with $A$
being at the origin and $B$ far away from the real axis. It suffices
to consider free boundary conditions on the real axis, since the work
of \cite{Dubail10} has already demonstrated that fixed or mixed
boundary conditions on the positive real axis can be accommodated by
fusing the operator that inserts the propagating (spin and FK)
clusters with an appropriate boundary condition changing operator at
the origin.

The results for the critical exponents can be stated in terms of
the Kac parametrisation of CFT
\be
 h_{r,s} = \frac{\left( r-s \, \kappa/4 \right)^2 -
 \left( 1 - \kappa/4 \right)^2}{\kappa} \,,
 \label{Kac}
\ee
where $2 \le \kappa \le 4$ parameterises the number of states $Q \in
[0,4]$ via
\be
 Q = 4\left( \cos \frac{\kappa \,\pi}{4} \right)^2 \,.
\ee
In the bulk case (geometry of the plane) the critical exponent
corresponding to the two-point function defined above is denoted $h(Q)
= h(Q; \sigma_1^{\alpha_1} \sigma_2^{\alpha_2} \ldots
\sigma_{\ell}^{\alpha_\ell})$, and the correlation function (or
probability) of the prescribed event decays like $P \propto x^{-4
  h(Q)}$ for $x \gg 1$. Equivalently, on a long cylinder of size $L
\times M$ with $M \gg L$, periodic boundary conditions in the
$L$-direction, and $A$ and $B$ identified with the opposite ends of
the cylinder, the decay is exponential: $P \propto {\rm e}^{- 4 \pi
  (M/L) \, h(Q)}$. In the boundary case (geometry of the half plane)
the critical exponent for free boundary conditions is denoted
$\widetilde{h}(Q)$.

We can write the bulk and boundary exponents in a unified way as
\begin{eqnarray}
 h(Q) &=& h_{q_1,q_2} \,, \label{conj_bulk} \\
 \widetilde{h}(Q) &=& h_{1+2q_1,1+2q_2} \label{conj_boundary} \,,
\end{eqnarray}
where the right-hand side refers to (\ref{Kac}) and we have defined a
``charge'' $(q_1,q_2)$. The result of \cite{DS_NPB87} is then that
$\ell$ propagating FK clusters correspond to
\be
 (q_1,q_2) = (0,\ell) \,, \qquad \qquad \ \ \  \mbox{for FK clusters.}
 \label{res_FK}
\ee
In the case of spin clusters with $\ell_1$ thin DW and $\ell_2$ thick
DW the result of \cite{Dubail10} is that
\be
 (q_1,q_2) = (\ell_1-\ell_2,2\ell_1) \,, \qquad \mbox{for spin clusters.}
 \label{res_spin}
\ee

These results strongly suggest that the costs for creating an
excitation consisting of several types of interfaces simply add up at
the level of the charge $(q_1,q_2)$. This is a well-known situation in
the Coulomb gas (CG) formalism. But note that whereas the result
(\ref{res_FK}) for FK clusters indeed admits a CG derivation
\cite{DS_NPB87} no such argument has yet been established for the spin
cluster result (\ref{res_spin}). Admitting the additivity of charges,
we can nevertheless read off the following rules from (\ref{res_spin})
\be
 (q_1,q_2) = \left \lbrace
 \begin{array}{ll}
  (1,2) \quad & \mbox{for a thin DW} \,, \\
  (-1,0)      & \mbox{for a thick DW} \,.
 \end{array} \right.
 \label{rules_spin}
\ee

We now argue that two additional rules are needed for dealing with the
insertion of FK clusters inside a given spin cluster. Indeed, when a
single FK cluster is inserted ($\alpha_i = 1$) we create a pair of
interfaces separating the hull of the FK cluster from the hull of the
surrounding spin cluster. The insertion of each subsequent FK cluster
(when $\alpha_i > 1$) will in addition create an interface separating
the FK cluster from the one preceeding it. To match (\ref{res_FK}) we
clearly need $(q_1,q_2) = (0,1)$ in the latter case.  To deal with the
former, we note that using our conventions, a configuration with just
$\ell$ FK clusters can either be labelled $1^\ell$ (if each FK cluster
has the same colour) or $1^1 2^1 \ldots \ell^1$ (if each FK cluster
has a different colour). In either case, additivity of charges and
consistency with (\ref{res_FK}) requires the correct rule to be
\be
 (q_1,q_2) = \left \lbrace
 \begin{array}{ll}
  (-1,-1) \quad & \mbox{for the first FK in a given spin cluster} \,, \\
  (0,1)         & \mbox{for each subsequent FK in that cluster} \,.
 \end{array} \right.
 \label{rules_FK}
\ee

The main result of this paper is the following conjecture:
\begin{itemize}
\item Consider the excitation labelled as in (\ref{gen_label}).
  Compute the corresponding charge $(q_1,q_2)$ by adding up the
  contributions from the $\sigma_k$, using rules (\ref{rules_spin}),
  and those of the $\alpha_k$, using rules (\ref{rules_FK}). Then the
  bulk and boundary critical exponents are given by
  (\ref{conj_bulk})--(\ref{conj_boundary}).
\end{itemize}

\subsection{Remarks on the first spin cluster}

The reader will have noticed that the nature (thin or thick) of the DW
are determined by the relative colours of two spin clusters. It
therefore remains to discuss carefully how to account for the first
one of the spin clusters making up the excitation.

In the bulk case, when only a single spin cluster is present
($\ell=1$), it will almost surely wrap around the periodic direction
(i.e., separate the neighbourhoods $A$ and $B$). Our results do {\em
  not} apply to this case. Indeed, it is well known that the critical
exponent in this case is the magnetic exponent $h_{1/2,0}$ (resp.\
$h_{0,\alpha_1}$) for a spin cluster with $\alpha_1=0$ (resp.\
$\alpha_1 \ge 1$) propagating FK clusters, cf.~(\ref{res_FK}). If, on
the other hand, we impose the additional constraint that the spin
cluster must not wrap around the periodic direction, our results {\em
  do} apply. The constraint then amounts to the presence of one {\em
  thick} DW that separates the spin cluster from itself.

In the boundary case, our results always apply, provided that the
first spin cluster is counted as a {\em thin} DW. This makes intuitive
sense, since the free boundary conditions permit the spin cluster to
touch the boundary, just as in the case of two spins clusters with
different colours.

\subsection{Possible values of the exponents $h_{r,s}$}
\label{sec:poss_val}

It is obvious that all excitations of the type (\ref{gen_label}) will lead
to charges $(q_1,q_2)$, where $q_1$ and $q_2$ take integer (positive
{\em or} negative) values. We now wish to clarify precisely which
values $(q_1,q_2) \in \mathbb{Z}^2$ can be obtained through the
joint spin-FK observables.

The spin part of the excitation is described by the numbers $(\ell_1,\ell_2)$
of thin and thick DWs, as in (\ref{res_spin}). We first discuss the boundary
case, for which the allowed values are $\ell_1 \ge 1$ (since the first DW
is thin) and $\ell_2 \ge 0$. Using rules (\ref{rules_spin})--(\ref{rules_FK})
one can then deduce the possible values of the charge $(q_1,q_2)$:
\begin{itemize}
\item For $q_1 \ge 1$, we have the constraint $q_2 \ge 2 q_1$. Indeed,
the excitation with $(\ell_1,\ell_2) = (q_1,0)$ and no FK clusters gives
the charge $(q_1,2 q_1)$, while the one with $(\ell_1,\ell_2) = (q_1+1,0)$
and a single FK cluster on one of the spin clusters gives the charge
$(q_1,2q_1+1)$. Any higher value of $q_2$ can be attained by placing
additional FK clusters on the same spin cluster.
\item For even $q_1 \le 0$, we have the constraint $q_2 \ge \frac{q_1+2}{2}$.
Indeed, the excitation with $(\ell_1,\ell_2) = (1,-\frac{q_1}{2})$ and one FK cluster
on each one of the spin clusters corresponds to $q_2 = \frac{q_1+2}{2}$,
while any higher value can be obtained by adding further FK clusters.
\item For odd $q_1 \le 0$, we have the constraint $q_2 \ge \frac{q_1+3}{2}$.
Indeed, the excitation with $(\ell_1,\ell_2) = (1,-\frac{q_1-1}{2})$ and one FK
cluster on each of the spin clusters except the first one saturates the inequality
on $q_2$, whose value can be increased by the addition of more FK clusters.
\end{itemize}

The discussion of the bulk case is similar, except that the allowed values of
$(\ell_1,\ell_2)$ are $\ell_1,\ell_2 \ge 0$ with $\ell_1 \neq 1$ (because of
the periodic boundary conditions), and of course $(\ell_1,\ell_2) \neq (0,0)$ in order
to have a non-trivial excitation. The leads only to a very minor modification
of the constraints on the charge $(q_1,q_2)$ derived in the boundary case:
For $q_1 < 0$ we have now $q_2 \ge \lceil \frac{q_1}{2} \rceil$.

In conclusion, the spin-FK observables described in this paper make it
possible to produce (say, in the bulk case) all the Kac table exponents
$h_{r,s}$ with integer indices $(r,s)$ above or below both of the lines
$s=2r$ and $s=r/2$.%
\footnote{That is, up to discretisation effects due to the fact that $(r,s)$ are
integers---see above for precise statements.}
The two cones in between these lines are not accessible by the spin-FK observables.

\section{Transfer matrix construction}
\label{sec:TM}

We now describe a transfer matrix whose spectrum contains all of the
excitations discussed in section~\ref{sec:obs}.

First recall that a {\em partition} ${\cal P}$ of a set $X$ is a set
of nonempty subsets of $X$ such that every element $i \in X$ is in
exactly one of these subsets. The elements of ${\cal P}$ (i.e., the
nonempty subset of $X$) are called {\em blocks}. A block containing
precisely one element of $X$ is called a {\em singleton}. A partition
${\cal P}_a$ is said to be a {\em refinement} of ${\cal P}_b$, and we
write ${\cal P}_a \preceq {\cal P}_b$, provided that each block in
${\cal P}_a$ is a subset of some block in ${\cal P}_b$. This defines a
partial order of the partitions of $X$.

We shall need a few simple operators acting on ${\cal P}$. We denote
the identity operator by $I$. For $i,j \in X$ the {\em join operator}
$J_{ij}$ acts as $I$ if $i$ and $j$ belong to the same block; if not,
it amalgamates the block containing $i$ with the block containing $j$
so as to form a single block. The {\em detach operator} $D_i$ acts by
detaching $i$ from its block, i.e., by transforming it into a new singleton
block $\{i\}$.

If we associate a Potts spin $\sigma_i$ with each $i \in X$ we can
finally define the {\em indicator operator} as $\Delta_{ij} =
\delta_{\sigma_i,\sigma_j} \cdot I$.

In the representation theory of the Temperley-Lieb algebra, $D_i$ is
usually defined in a slightly different manner. Namely, $D_i$ acts as
$\widetilde{Q} \cdot I$ if $i$ is a singleton (i.e., it gives a Boltzmann weight
$\widetilde{Q}$); otherwise it transforms $i$ into a singleton (with a
Boltzmann weight $1$). Then, setting
$e_{2i-1} = \widetilde{Q}^{-1/2} D_i$ and $e_{2i} = \widetilde{Q}^{1/2} J_{i,i+1}$,
the $e_k$ provide a representation of the Temperley-Lieb algebra \cite{Martin}
corresponding to the $\widetilde{Q}$-state Potts model. In our construction
of the transfer matrix we shall account for $Q$ in a different manner, which
is why we have set $\widetilde{Q}=1$ above.

\subsection{States and elementary transfer matrices}

The states of the Potts model in the spin-FK representation are
given by a triplet of partitions $({\cal P}_1,{\cal P}_2,{\cal P}_3)$
of the set $X$ of vertices within a row of the lattice ${\cal L}$.

The first partition ${\cal P}_1$ is arbitrary (not necessarily planar)
and describes the spin colours: we have $\sigma_i = \sigma_j$ if and
only if $i,j$ belong to the same block in ${\cal P}_1$.  The second
partition ${\cal P}_2$ is planar and describes the connectivity of
spin clusters. We have ${\cal P}_2 \preceq {\cal P}_1$, since spins in
the same spin cluster have equal colours. The third partition ${\cal
  P}_3$ is again planar and describes the connectivity of FK
clusters. We have ${\cal P}_3 \preceq {\cal P}_2$, since spins in the
same FK cluster are also in the same spin cluster.

We represent graphically the transfer matrix as
\be
 \vspace*{1.5cm}\includegraphics[height=1.5cm]{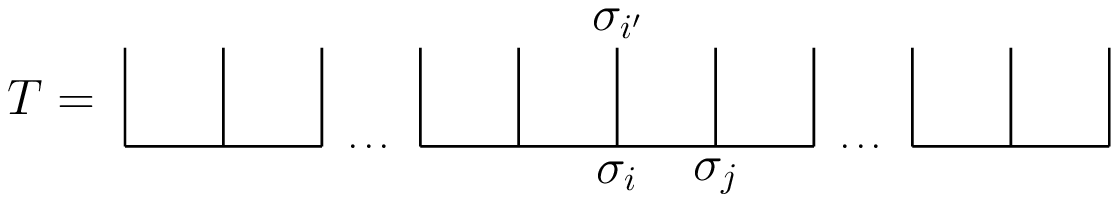}
 \label{Tpic}
\ee
The elementary transfer matrix $T_{\rm h}$ that adds a horizontal edge between
vertices $i$ and $j$ acts in ${\cal P}_1 \otimes {\cal P}_2 \otimes
{\cal P}_3$ as follows:
\begin{equation}
 T_{\rm h} = 1 \cdot (I-\Delta_{ij}) \otimes I \otimes I +
 ({\rm e}^K - 1) \cdot \Delta_{ij} \otimes J_{ij} \otimes J_{ij} +
 1 \cdot \Delta_{ij} \otimes J_{ij} \otimes I \,,
 \label{Th}
\end{equation}
where the number before the dot ($\cdot$) is the Boltzmann weight.
The first term corresponds to the two spins $\sigma_i$ and $\sigma_j$
being different, in which case $i,j$ can neither be in the same spin
cluster, nor in the same FK cluster. The second term corresponds to
$i,j$ being in the same FK cluster, hence also in the same spin
cluster.  Finally, the third term corresponds to $i,j$ having the same
spin without being in the same FK cluster; in this case the spin
clusters must be joined up since $i,j$ are neighbours on the lattice.

We can similarly define the elementary transfer matrix $T_{\rm v}$
that adds a vertical edge edge to the lattice, by propagating the
vertex $i$ to a new vertex $i'$.  It acts in ${\cal P}_1 \otimes {\cal
  P}_2 \otimes {\cal P}_3$ like
\begin{equation}
 T_{\rm v} = 1 \cdot (I-\Delta_{ii'}) \otimes D_i \otimes D_i +
 ({\rm e}^K - 1) \cdot \Delta_{ii'} \otimes I \otimes I +
 1 \cdot \Delta_{ii'} \otimes I \otimes D_i \,,
 \label{Tv}
\end{equation}
and each term has the same interpretation as in $T_{\rm h}$. Since $T$
builds up the partition function $Z$, it must also account for the summation
$\sum_{\sigma}$ over the spin varibles, {\it cf.}~(\ref{Potts}). The easiest 
convention is to let the action of $T_{\rm v}$ be accompanied by a sum over 
$\sigma_{i'}$. This sum is dealt with in the same way as in
\cite{Dubail10}. Namely, let $\Sigma$ denote the set of spin values
$\{\sigma_i\}$ being used in a given state, so that the number of
elements $|\Sigma|$ is just the number of blocks in ${\cal
  P}_1$. After summing over $\sigma_{i'}$, the second and third terms
in (\ref{Tv}) each correspond to a single non-zero contribution.  The
first term gives $|\Sigma|-1$ contributions where $\sigma_{i'} \neq
\sigma_i$, but $\sigma_{i'} \in \Sigma$. The remaining contributions,
where $\sigma_{i'} \neq \sigma_i$ and $\sigma_{i'} \notin \Sigma$ can
be regrouped (using the overall $S_Q$ symmetry) as a single
contribution, with Boltzmann weight $Q-|\Sigma|$, in which $\{i'\}$
becomes a singleton in ${\cal P}'_1$. It is precisely because of this
regrouping that it now makes sense to promote $Q$ to an arbitrary 
real variable \cite{Dubail10}.

The complete row-to-row transfer matrix $T$ for a system of size $L$,
shown graphically in~(\ref{Tpic}),
is then obtained as the product of the elementary transfer
matrices corresponding to all horizontal and vertical edges in a row:
\be
 T = \left( \prod_{i=1}^L T_{\rm v}^{(i)} \right) \times
     \left( \prod_{i=1}^{L'} T_{\rm h}^{(i,i+1)} \right) \,,
\ee
where $L' = L-1$ in the boundary case (strip geometry) and $L'=L$ with
indices $i$ considered modulo $L$ in the bulk case (cylinder geometry).

Several remarks are in order:
\begin{enumerate}
\item When representing ${\cal P}_1$ one can replace the actual spin
  values $\sigma_i$ by colour labels $c_i$ defined such that $c_i =
  c_j$ if and only if $\sigma_i = \sigma_j$.  Using the $S_Q$ symmetry
  these colour labels can be brought into a standard form, thus
  reducing the number of basis states. Details of this construction
  are given in \cite{Dubail10}.
\item For a system of size $L$, at most $L$ different colour labels
  are used. Therefore the number of basis states depends only on $L$,
  and not on $Q$. In particular the state space is finite.
\item The three partition algebras ${\cal P}_1$, ${\cal P}_2$ and
  ${\cal P}_3$ are coupled by virtue of (\ref{Th})--(\ref{Tv}). It is
  an interesting problem, beyond the scope of this paper, to analyse
  in details this situation from a representation theoretical point of
  view.
\item If one neglects the information on FK clusters contained in
  ${\cal P}_3$, the second and third terms in (\ref{Th})--(\ref{Tv})
  can be resummed, and one recovers the construction of
  \cite{Dubail10}.
\item If one neglects the information on spin clusters contained in
  ${\cal P}_2$, the first and third terms in (\ref{Th})--(\ref{Tv}) can be
  resummed. The result is then independent of ${\cal P}_1$ and reproduces the
  well-known Temperley-Lieb representation \cite{Martin}.
\item In the above construction we have accounted for $Q$ in the partition
  algebra ${\cal P}_1$. Therefore, the detach operators $D_i$ acting on
  ${\cal P}_2$ and ${\cal P}_3$ have parameter $\widetilde{Q} = 1$. In
  particular, the FK clusters described by ${\cal P}_3$ can be thought of
  as percolation ($\widetilde{Q}=1$) clusters inside the spin clusters
  described by ${\cal P}_1$ and ${\cal P}_2$. Leaving $\widetilde{Q}$ as
  a free parameter would amount to studying FK clusters of a second
  $\widetilde{Q}$-state Potts model defined on top of the spin clusters of
  the original $Q$-state Potts model. This appears to be an interesting
  way of coupling a pair of $(Q,\widetilde{Q})$-state Potts models---and
  we intend to report further on this elsewhere.
\end{enumerate}

The leading eigenvalue $\Lambda_0$ of the transfer matrix $T$ gives
the ground state free energy $f_0 = -\frac{1}{L} \log \Lambda_0$. This
$f_0$ coincides precisely with that of the usual FK transfer matrix
\cite{Blote82}, even when $Q$ is non-integer. Its finite-size
corrections possess a universal $L^{-2}$ term whose coefficient
determines the central charge of the corresponding CFT
\cite{CardyBloteNightingale}.

\subsection{Correlation functions}

To obtain the two-point correlation defined in section~\ref{sec:obs}
we need a variant transfer matrix $T'$ which imposes the propagation
of the defect labelled as in (\ref{gen_label}) along the (imaginary) time
direction of the cylinder (or strip). From its leading eigenvalue
$\Lambda'_0$ one can determine the energy gap $\Delta f = -\frac{1}{L}
\log(\Lambda'_0 / \Lambda_0)$ whose finite-size scaling in turn
determines the critical exponents $h(Q)$ and $\widetilde{h}(Q)$
\cite{CardyBloteNightingale,JJreview}.

To construct $T'$ we modify the basis states ${\cal P}_1 \otimes {\cal
  P}_2 \otimes {\cal P}_3$ by {\em marking} some of the blocks in the
partitions ${\cal P}_2$ and ${\cal P}_3$. First, we mark $\ell$ blocks
in ${\cal P}_2$ whose spin colours in ${\cal P}_1$ coincide with the
choice of $\{\sigma_1,\sigma_2,\ldots,\sigma_\ell\}$ in
(\ref{gen_label}). Second, let ${\cal P}_3$ be a refinement of ${\cal
  P}_2$ such that the $k$'th marked block in ${\cal P}_2$ is refined
into at least $\alpha_k$ blocks in ${\cal P}_3$, of which precisely
$\alpha_k$ are marked.

In order to conserve the marked clusters in the transfer matrix
evolution, none of the marked clusters must be ``left behind'', and
two distinct marked clusters must not be allowed to link up.  Imposing
these rules in a precise way is tantamount to defining a modified
action of the join and detach operators, $J_{ij}$ and $D_i$, on the
marked basis states. Namely, $J_{ij}$ is modified to give a zero
Boltzmann weight if $i,j$ are in distinct marked blocks (this prevents
a marked block from disappearing), and $D_i$ is modified to give a
zero Boltzmann weight if $\{i\}$ is a marked singleton block (this
prevents marked blocks from being ``left behind'').  Finally, when
$J_{ij}$ joins a marked and an unmarked block, the result is a
marked block.

The full state space corresponding to the excitation (\ref{gen_label})
is generated by letting $T'$ act a sufficient number of times on a
suitable initial basis state. This initial state is such that the
$k$'th marked block in ${\cal P}_2$ consists of $\alpha_k$ consecutive
points $\{i_k,i_k+1,\ldots,i_k+\alpha_k-1\}$ in a row of the
lattice. In the refining partition ${\cal P}_3$ each of these
$\alpha_k$ points is a marked singleton. Finally, those of the $L$
points which are not marked in this construction of the initial state
are taken as singletons, both in ${\cal P}_2$ and in ${\cal P}_3$.

In summary, the modified transfer matrix $T'$ keeps enough
information, both about the mutual colouring of the sites and about
the connectivity of spin and FK clusters, to give the correct
Boltzmann weights to the different configurations, even for
non-integer $Q$, and to follow the time evolution of the excitation
defined by (\ref{gen_label}).

\section{Exact diagonalisation results}
\label{sec:diag}

We have numerically diagonalised the transfer matrix in the spin-DW
representation for cylinders (resp.\ strips) of width up to $L=8$
(resp.\ $L=9$) spins. We
verified that the leading eigenvalue $\Lambda_0$ in the ground state
sector coincides with that of the FK transfer matrix, including for
non-integer $Q$.  As to the excitations $\Lambda'_0$, we explored
systematically all possible colouring combinations (\ref{gen_label})
for up to $\ell=3$ marked spin clusters with different choices for the
number of FK clusters $\alpha_k$, for a variety of values of the
parameter $\kappa$.

Finite-size approximations of the critical exponents $h(L)$ and
$\widetilde{h}(L)$ were extracted from the leading eigenvalue in each
sector, using standard CFT results
\cite{CardyBloteNightingale,JJreview}, and fitting both for the
universal corrections in $L^{-2}$ and the non-universal $L^{-4}$ term.
These approximations were further extrapolated to the $L \to \infty$
limit by fitting them to first and second order polynomials in
$L^{-1}$, gradually excluding data points corresponding to the
smallest $L$. Error bars were obtained by carefully comparing the
consistency of the various extrapolations.

\begin{table}
\begin{center}
\begin{tabular}{r|rrrr}
  $p=\frac{\kappa}{4-\kappa}$ &  $1^1 2^0$ & $1^2 2^0$ & $1^1 1^0$ & $1^1 2^0 3^0$ \\ \hline
  $2$ &  3.0000(1) &  5.01(1) &  ---     &  3.99(1) \\
  $3$ &  5.005(5)  &  8.05(5) &  9.03(2) &  7.03(3) \\
  $4$ &  6.97(2)   & 11.02(3) & 11.2(1)  &  9.95(5) \\
  $5$ &  8.85(10)  & 13.9(1)  & 13.4(3)  & 12.7(3)  \\[0.1cm]
Exact & $2p-1$ & $3p-1$ & $2p+3$ & $3p-2$ \\
\end{tabular}
\end{center}
\caption{\label{tab1}
  Bulk critical exponents corresponding to four different sector
  labels (\ref{gen_label}), as functions of the parameter
  $p = \frac{\kappa}{4-\kappa}$.
  The conjectured exponents read
  $h_{1,3}$ for sector $1^1 2^0$,
  $h_{1,4}$ for sector $1^2 2^0$,
  $h_{3,1}$ for sector $1^1 1^0$,
  and $h_{2,5}$ for sector $1^1 2^0 3^0$.
  The table entries
  give the value of $|\rho|$, when (\ref{Kac}) is rewritten as
  $h_{r,s} = (\rho^2-1)/(4p(p+1))$, with error bars shown in parentheses.} 
\end{table}

Representative final results for the bulk case (periodic boundary
conditions) are shown in Table~\ref{tab1}.  Corresponding results for
the boundary case (free boundary conditions) are given in
Table~\ref{tab2}. In all cases the agreement with the conjecture
made in section~\ref{sec:obs} is very good.

\begin{table}
\begin{center}
\begin{tabular}{r|rrrr}
  $p=\frac{\kappa}{4-\kappa}$ &  $1^1 2^0$ & $1^2 2^0$ & $1^1 1^0$ & $1^1 2^0 3^0$ \\ \hline
  $2$ &  5.001(2) &  9.00(1) &  9.01(2)  &  7.002(4) \\
  $3$ &  9.01(1)  & 15.0(1)  & 13.02(2)  & 13.05(8)  \\
  $4$ & 12.92(5)  & 20.98(2) & 17.02(3)  & 18.85(10) \\
  $5$ & 16.7(2)   & 26.7(2)  & 21.05(10) & 24.5(3)   \\[0.1cm]
Exact & $4p-3$    & $6p-3$  & $4p+1$   & $6p-5$   \\
\end{tabular}
\end{center}
\caption{\label{tab2}
  Boundary critical exponents with free boundary conditions, corresponding to
  four different sector
  labels (\ref{gen_label}), as functions of the parameter
  $p = \frac{\kappa}{4-\kappa}$.
  The conjectured exponents read
  $h_{3,7}$ for sector $1^1 2^0$,
  $h_{3,9}$ for sector $1^2 2^0$,
  $h_{-1,3}$ for sector $1^1 1^0$,
  and $h_{5,11}$ for sector $1^1 2^0 3^0$.
  The table entries
  give the value of $|\rho|$, when (\ref{Kac}) is rewritten as
  $h_{r,s} = (\rho^2-1)/(4p(p+1))$, with error bars shown in parentheses.} 
\end{table}

\begin{figure}[ht]
\centering
\subfigure[]{
\includegraphics[width=4.8cm]{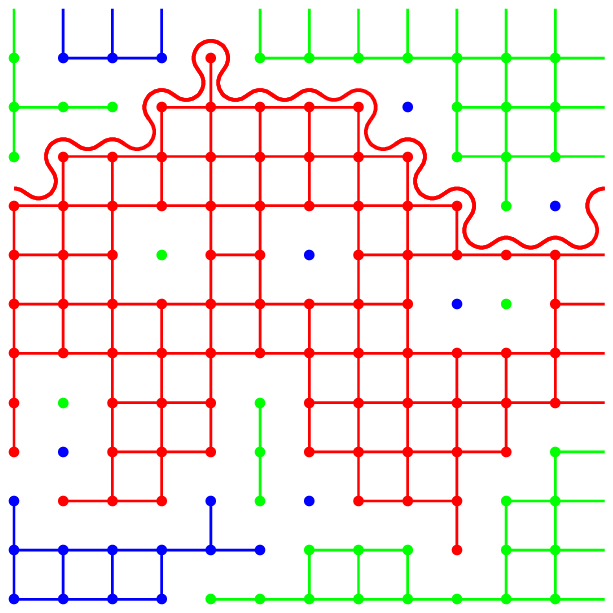}
\label{fig2a}
}
\subfigure[]{
\includegraphics[width=4.8cm]{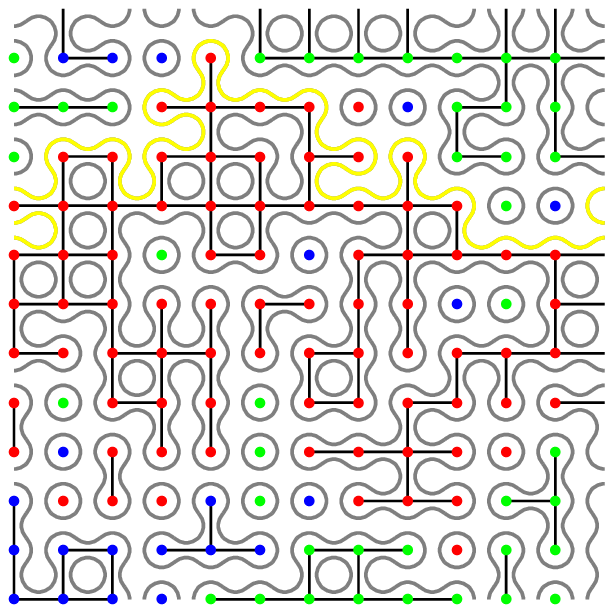}
\label{fig2b}
}
\subfigure[]{
\includegraphics[width=4.8cm]{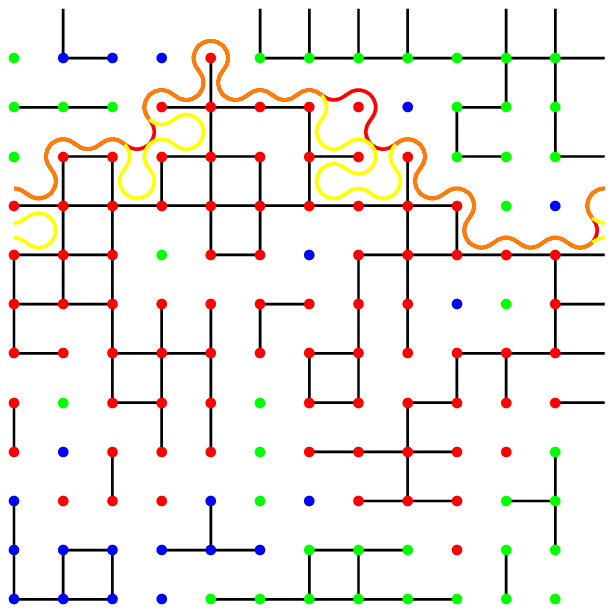}
\label{fig2c}
}
\label{fig2}
  \caption{Various critical curves in the $3$-state Potts model. 
  The three pictures show the same configuration with different 
  geometrical features emphasised. (a) Spin cluster configuration.
  There is a non-trivial ({\it i.e.} non-homotopic to a point) red 
  cluster wrapping along one periodic boundary condition, but not 
  the other one. One of 
  its boundaries is shown in red, and contributes to the fractal 
  dimension $d_{\rm spin}=2-2 h_{1,0}$. (b) Same spin configuration,
  but this time the FK clusters and their corresponding boundaries 
  are shown. The non-trivial red cluster is broken into
  smaller FK clusters, one of them remaining non-trivial.
  One of the two boundaries of the percolating
  FK cluster is shown in yellow and has a fractal 
  dimension $d_{\rm FK}=2-2 h_{0,1}$ in the continuum limit. (c) 
  Superposition of the previous red and yellow interfaces. Their intersection,
  shown in orange, corresponds to the set of points where the FK cluster
 touches the hull of its surrounding spin cluster. Its fractal dimension
 is $d_{2,1}=2-2 h_{2,1}$.}
\end{figure}

\section{Monte Carlo simulations}
\label{sec:MC}

Although we the result of section~\ref{sec:obs} apply to a general
excitation (\ref{gen_label}), only a few of the resulting scaling
dimensions $h_{r,s}$ are relevant (i.e., $0 \le h_{r,s} < 1$) for some
$Q$ in the interval $0 \le Q \le 4$. Moreover, some of the excitations
are relevant only in a part of the interval, where there are not
``enough colours'' to realise the choice of different $\sigma_k$ used
in (\ref{gen_label}). An example of this situation is the bulk
operator $1^1 2^0 3^0$ with scaling dimension $h_{2,5}$, which is
relevant only for $0 \le Q \le 2$, meaning that $Q$ is not large
enough to accommodate the three different spin colours defining the
excitation.

These remarks become important if we want to measure the fractal
dimension corresponding to a given excitation in a Monte Carlo
simulation. For convenience we restrict the discussion to bulk
critical exponents. Among the excitations which involve both spin and
FK degrees of freedom, it appears that only two are relevant and
physical (in the above sense). The first of these is $1^1$ with
wrapping around the periodic direction disallowed. It describes the
insertion of a spin cluster with one FK cluster inside it. The charge
is $(q_1,q_2) = (-1,0) + (-1,-1)$ from
(\ref{rules_spin})--(\ref{rules_FK}) leading to the scaling dimension
$h_{-2,-1} = h_{2,1}$ by application of (\ref{conj_bulk}). The corresponding
codimension
\be
 d_{2,1} = 2 - 2 h_{2,1} = 3 - \frac{6}{\kappa}
\ee
is thus the fractal dimension of the set of points where the FK cluster
touches the hull of its surrounding spin cluster. We have $0 \le
d_{2,1} \le 2$ for all $0 \le Q \le 4$. An example of a geometrical curve
corresponding to the dimension $d_{2,1}$ is shown in Fig.~\ref{fig2c}.

The second excitation of interest is $1^1 2^0$. The charge is now
$(q_1,q_2) = 2 \times (1,2) + (-1,-1)$ from
(\ref{rules_spin})--(\ref{rules_FK}), and the corresponding scaling
dimension reads $h_{1,3}$ by (\ref{conj_bulk}). The fractal dimension
\be
 d_{1,3} = 2 - 2 h_{1,3} = 4 - \kappa
\ee
describes the same set of points as above, but with the additional
constraint that a second spin cluster (the $2^0$ part of the label)
adjacent to the one whose hull is being touched by its internal FK
cluster (the $1^1$ part of the label) must now also propagate all the
way to infinity. Note that $d_{1,3}$ and $d_{2,1}$ should 
coincide for the Ising model ($Q=2$) and indeed we find $d_{1,3}=d_{2,1}=1$
in this case.

We checked numerically these fractal dimensions using Monte Carlo 
(MC) methods. Efficient MC algorithms for the non-integer $Q$-state Potts 
model are already available. We chose here to work with the Chayes--Machta
algorithm~\cite{Chayes_Machta} that works for $Q \in \left[ 1,4 \right]$. 
This algorithm is easy to implement and allows one to keep track of both FK and 
spin clusters, even for non-integer $Q$~\cite{Zatelepin_Shchur}. 

\begin{figure}[t]
\begin{center}
    \includegraphics[width=12.0cm]{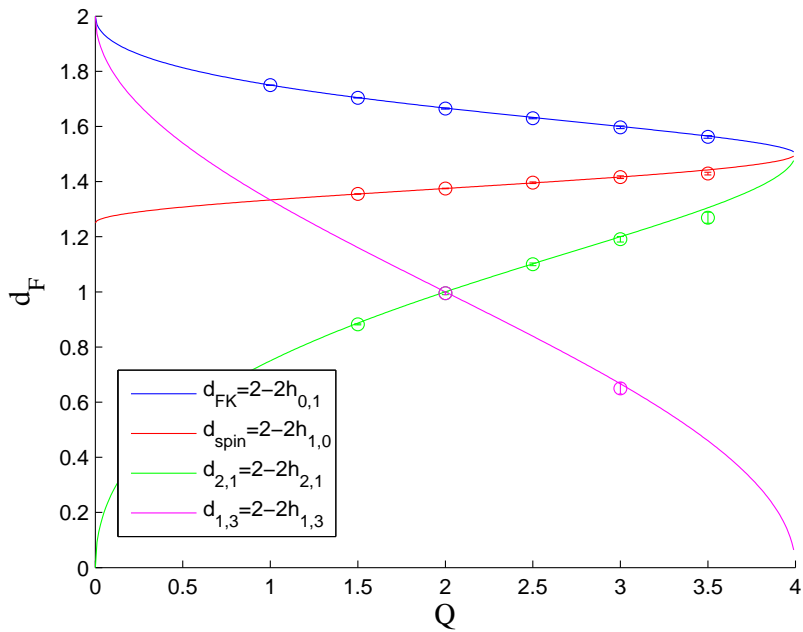}
  \caption{Fractal dimensions $d_{\rm FK}$, $d_{\rm spin}$, 
  $d_{2,1}$ and $d_{1,3}$ (see text for details) as 
  functions of $Q$. The small circles with their error bars are 
  obtained from Monte Carlo simulations, while solid lines represent the CFT
  prediction.}
  \label{fig3}
\end{center}
\end{figure}

The elementary step of the algorithm reads:
\begin{itemize}
\item Find all the FK clusters in the configuration.
\item Independently label the clusters as active or
inactive with respective probability $\frac{1}{Q}$
and $\frac{Q-1}{Q}$. Sites belonging to an active
clusters are said to be active.
\item Erase all bonds. Independently add bonds 
between pairs of active sites with probability $p_c=1-\mathrm{e}^{-K}=\frac{\sqrt{Q}}{1+\sqrt{Q}}$ .
The resulting clusters are the new FK clusters. 
\end{itemize} 
Active sites correspond to spins in a given colour $Q_0$, so that
the spin clusters of this colour can be obtained by performing the 
bond-adding step of Chayes--Machta at zero temperature, {\it i.e.} by 
replacing $p_c$ by $p_0=1$. The 
clusters and their boundaries are then detected in a standard way.

Since the fractal dimensions we wish to measure 
are bulk properties, we work with an $L \times L$ lattice with periodic boundary 
conditions. We perform the statistics on non-trivial 
clusters wrapping once around one of the periodic boundaries.
The fractal dimensions $d_{\rm F}$ are then obtained by measuring the mean 
value $\ell$ of the curve length as a function of $L=32, 64, \dots, 800$; using
the relation $\ell \sim A L^{d_{\rm F}}$.

This algorithm allowed us to recover as a check the well-known 
fractal dimensions of the interfaces of the FK and 
spin clusters, which read respectively $d_{\rm FK}=2-2 h_{0,1}$ 
and $d_{\rm spin}=2-2 h_{1,0}$. The dimension $d_{2,1}=2-2 h_{2,1}$
of the set of points where the FK cluster
touches the hull of its surrounding spin cluster can be measured
without much more complication for $Q>1$.
To evaluate the dimension $d_{1,3}=2-2 h_{1,3}$, we restrict
the set of points considered when measuring $d_{2,1}$ to the points
whose adjacent spin cluster also wraps around the periodic 
boundary condition. As the Chayes--Machta for non-integer $Q$
only keeps track of one spin cluster, we were able to measure 
$d_{1,3}$ only for $Q$ integer. Note also that $d_{1,3}$
makes sense physically only for $Q \geq 2$.

Results are shown in Fig.~\ref{fig3}. The agreement with our predictions 
is very good. The small deviations we observe when $Q$ becomes close
to $Q=4$ are expected, as logarithmic corrections are known to occur
in this case.

\section{Conclusion}
\label{sec:disc}

We have defined geometrical observables that keep track of both FK and spin
clusters for any real $Q \in \left[ 0,4\right]$. We conjecture that such
observables are conformally invariant and we provide exact formulas for the bulk
and boundary critical exponents. Our results are supported by extensive 
transfer matrix and Monte Carlo computations.
It is quite remarkable that all the exponents are of the Kac form $h_{r,s}$;
an analytical understanding of those results would probably shed some light on this
rather intriging point.
We note also that we now have enough observables to cover all the Kac table $h_{r,s}$
for any integer choice of $(r,s)$, except in the two cones delimited by the straight
lines $s=2r$ and $s=r/2$ (see section~\ref{sec:poss_val}).
This remark is particulary important in
the context of Logarithmic CFT (LCFT), where including more involved geometrical 
observables in the theory might yield some possibly unknown, interesting
logarithmic features. In particular, our transfer matrix $T$ at logarithmic points 
should have a much more complicated structure that the ones considered so far in lattice
regularisations of LCFTs (see {\it e.g.} \cite{Vasseur1,Vasseur2}).

\section*{Acknowledgments}

We thank Hubert Saleur for stimulating discussions and collaboration on
related work.
This work was supported by the Agence Nationale de la Recherche (grant
ANR-10-BLAN-0414: DIME).

\end{document}